\documentclass[11pt]{article}

\def\tr{\;{\rm tr}\;}

\def\bra{\langle}   \def\ket{\rangle}
\def\pr{\prime}
\def\implies{\Rightarrow}

\newcommand{\dd}[2]{\frac {\partial #1}{\partial #2}}
\newcommand{\pdr}{\partial}

\newcommand{\beq}{\begin{eqnarray}}
\newcommand{\eeq}{\end{eqnarray}}

\newcommand{\half}{\frac{1}{2}}

\newcommand{\ov}[1]{\frac{1}{#1}}

\def\a{\alpha}      \def\b{\beta}   \def\g{\gamma}       
\def\d{\delta}      \def\D{\Delta}   
          \def\la{\lambda}      \def\La{\Lambda}
             \def\F{\Phi}

\newcommand{\N}{{1 \over N}}

\newcommand{\Ntr}{{{\rm tr} \over N}}

\newtheorem{sutra}{}

\newtheorem{bhashya}{}[sutra]

\input{epsf}

\begin{document}

%------------------------------------------------

\begin{titlepage}

\title{\normalsize \hfill {\tt arXiv:0708.3056 [hep-th]}\\ \vskip 5mm  \Large\bf
Schwinger-Dyson operator of Yang-Mills matrix models with ghosts and derivations of the graded shuffle algebra}

\author{Govind S. Krishnaswami}
\date{\normalsize Institute for Theoretical Physics \& Spinoza Institute \\
Utrecht University, Postbus 80.195, 3508 TD, Utrecht, The Netherlands \vspace{.2in} \\
Department of Mathematical Sciences \& Centre for Particle Theory, \\
Durham University, Science Site, South Road, Durham, DH1 3LE, UK \vspace{.2in}\\
Chennai Mathematical Institute, \\
Padur PO, Siruseri 603103, India.
\smallskip \\ e-mail: \tt govind.krishnaswami@durham.ac.uk \\ Nov 25, 2007}

\maketitle

\begin{quotation} \noindent {\large\bf Abstract } \medskip \\

We consider large-$N$ multi-matrix models whose action closely mimics that of Yang-Mills theory, including gauge-fixing and ghost terms. We show that the factorized Schwinger-Dyson loop equations, expressed in terms of the generating series of gluon and ghost correlations $G(\xi)$, are quadratic equations ${\cal S}^i G = G \xi^i G$ in concatenation of correlations. The Schwinger-Dyson operator ${\cal S}^i$ is built from the left annihilation operator, which does not satisfy the Leibnitz rule with respect to concatenation. So the loop equations are not differential equations. We show that left annihilation is a derivation of the graded shuffle product of gluon and ghost correlations. The shuffle product is the point-wise product of Wilson loops, expressed in terms of correlations. So in the limit where concatenation is approximated by shuffle products, the loop equations become differential equations. Remarkably, the Schwinger-Dyson operator as a whole is also a derivation of the graded shuffle product. This allows us to turn the loop equations into linear equations for the shuffle reciprocal, which might serve as a starting point for an approximation scheme.

\end{quotation}

\vfill \flushleft


% (gauge field theories, $1/N$, rings and algebras, )


Journal Reference: J. Phys. A: Math. Theor. 41 (2008) 145402

\thispagestyle{empty}

\end{titlepage}

\eject
%--------------------------------------------------

{\scriptsize \tableofcontents}

\normalsize

%--------------------------------
\section{Introduction}
\label{s-intro}
%--------------------------------

QCD, a quantum Yang-Mills theory is the best candidate for a theory of strongly interacting subatomic particles. It is an outstanding challenge to understand its non-perturbative features and their mathematical formulation. Yang-Mills theory includes as dynamical degrees of freedom, a collection of $N \times N$ matrices (gluons), where $N=3$ in nature. The limit as the number of colours $N \to \infty$ is a promising starting point to study this theory \cite{thooft-large-N}. Large-$N$ matrix models \cite{bipz} are simplified models for the dynamics of gluons.

In this paper, we establish some properties of a class of large-$N$ multi-matrix models that may be regarded as toy-models for gauge-fixed Yang-Mills theory. These models have both hermitian complex matrices (bosonic gluons) and hermitian grassmann matrices (fermionic ghosts) and we call them Yang-Mills matrix models with ghosts. However, these models are generically not supersymmetric. Our work is inspired by that of Makeenko and Migdal \cite{makeenko-migdal-eqn,migdal-phys-rpts,makeenko-book} on the loop equations. Matrix models and their loop equations may be formulated in several different ways \cite{staudacher,tavares,douglas-li,entropy-var-ppl,kazakov-marshakov,eynard,deform-prod-der}. There is a large literature on matrix models involving both bosonic and fermionic degrees of freedom, especially in the context of the matrix approach to M-theory \cite{bfss,ikkt,wati-taylor} and in the matrix regularisation of the supermembrane \cite{supermembrane}.

`Solving' a matrix-model can be regarded as determining the gluon and ghost correlation functions. We obtain quantum corrected equations of motion (large-$N$ factorised Schwinger-Dyson or loop equations) for these correlations. They involve a `classical' term linear in correlations, coming from the variation of the action. This is the `Schwinger-Dyson operator' ${\cal S}^i$ acting on correlations. ${\cal S}^i$ is built from the left annihilation operator. The loop equations also involve a quadratic `quantum' term in correlations, coming from the change in path integral measure. This involves concatenation products of gluon and ghost correlation tensors. However, the left annihilation operator does not satisfy the Leibnitz rule (i.e. is not a derivation) with respect to concatenation. So the loop equations are {\em not} differential equations. This is perhaps part of the reason why the loop equations have been difficult to solve, though they were derived for the Wilson loops of Yang Mills theory over 25 years ago. It is therefore imperative to uncover any hidden mathematical structures of the loop equations, which may help in solving them and placing them in their natural mathematical context.

The great success of calculus in solving problems of classical mechanics is due to the fact that the equations are differential equations, rather than, say, difference equations. So it is interesting to know whether there is some limit or approximation where the loop equations become differential equations. The main result of this paper is that this is indeed the case, in the limit where concatenation of correlations is replaced by their shuffle products. While concatenation arises from concatenation of loops, shuffle arises from the point-wise product of Wilson loops. We show that this picture is robust, in the sense that it is not spoiled by the inclusion of gauge fixing and ghost terms in the action. More precisely, we show that left annihilation as well as the Schwinger-Dyson operator ${\cal S}^i$, satisfy the Leibnitz rule with respect to the (graded) shuffle product of gluon and ghost correlations. The latter allows us to reduce the non-linear loop equations to linear equations for the shuffle-reciprocal of the generating series for correlations. Though this is not a property shared by generic matrix models, it does carry over to $3+1d$ gauge-fixed Yang-Mills theory. For, all we use is the algebraic structure of the action, and general properties of the large-$N$ limit.

In a previous paper \cite{deform-prod-der}, we proposed an approximation scheme to compute correlation functions by solving the loop equations, in the context of bosonic matrix models. We expanded the concatenation product around the shuffle product and used the shuffle reciprocal to reduce the loop equations to linear differential equations in the shuffle algebra at the $0^{\rm th}$ order of the expansion. In simple cases, this was shown to give a rough approximation to the exact correlations. Aside from this practical application, a mathematical lesson from our work is that it may be fruitful to view the loop equations as living in the differential bi-algebra formed by shuffle, concatenation and their derivations.

The property that ${\cal S}^i$ is a derivation of the shuffle product, is a finite and differential-algebraic reformulation in terms of correlation tensors, of a property of large-$N$ Yang-Mills theory mentioned in \cite{migdal-phys-rpts}. ${\cal S}^i$ is analogous to the path derivative of the area derivative operator, which was said to satisfy the Leibnitz rule with respect to point-wise products of Wilson loops. We hope that our alternative viewpoint and finite formulation of these infinite dimensional notions is useful to better understand the loop equations.

%The paper is organised as follows. In section \ref{s-include-ghost-matrices} we define the gluon and ghost correlations in the large-$N$ limit and their symmetries. The factorised Schwinger-Dyson loop equations for correlations are obtained in section \ref{s-sde-to-loop-eq}. They are expressed in terms of a left annihilation operator and the concatenation product (section \ref{s-LE-conc-left-ann}) and an example is provided in section \ref{s-ghost-mat-model}. The graded shuffle product is introduced in section \ref{s-graded-shuffle} and shown to be commutative for correlations of zero ghost number and to preserve the vanishing of ghost number. In section \ref{s-left-ann-derivation-graded-sh-prod} we show that left annihilation and some of its iterated commutators are derivations of the graded shuffle product of tensors of zero ghost number. In section \ref{s-var-of-action-is-derivation} we show that even after the inclusion of gauge fixing and ghost terms, the Schwinger-Dyson operator of a Yang-Mills matrix model with ghosts defines a derivation of the graded shuffle product. We end with a discussion in section \ref{s-discussion} where we provide a mathematical context for the loop equations and also suggest an application of our result to a possible approximation scheme to simplify the loop equations.

%---------------------------------------------
\section{Gluon-Ghost correlations in the large-$N$ limit}
\label{s-include-ghost-matrices}
%---------------------------------------------

Motivated by the Lagrangian of gauge-fixed Yang-Mills theory (say in a class of covariant gauges labeled by $\xi$)
    \beq
    {\cal L} &=& \tr\bigg\{\half \pdr_\mu A_\nu(\pdr^\mu A^\nu ~-~ \pdr^\nu A^\mu)
    ~-~ ig \pdr_\mu A_\nu [A^\mu,A^\nu] ~-~ {g^2 \over 4} [A_\mu,A_\nu][A^\mu,A^\nu]
    \cr && ~+~ \ov{2 \xi} (\pdr^\mu A_\mu)^2 ~+~ \pdr_\mu \bar c ~\pdr^\mu c
    ~-~  ig \pdr_\mu \bar c ~[A^\mu,c]  \bigg\},
    \label{e-gauge-fixed-ym-action}
    \eeq
we consider models with $\La$ matrices $A_i, ~~1 \leq i \leq \La$, which are $N \times N$ matrices in `colour' space. We will call them Yang-Mills matrix models with ghosts if their action is of the form
    \beq
    \tr S = \half \tr C^{ij} A_i A_j + \tr C^{ijk} A_i [A_j, A_k]
    -\ov{4 \a} \tr [A_i,A_j][A_k,A_l] g^{ik} g^{jl}.
    \label{e-action-ym-mat-mod-with-ghosts}
    \eeq
We call all matrices $A_i$ irrespective of whether they are gluons ($A_\mu$),
ghosts ($c$) or anti-ghosts ($\bar c$). Indices $i,j,k$ are shorthand for position
coordinates and polarisation indices. They also specify whether the
matrix is a gluon, a ghost or an anti-ghost via the ghost number of
an index
    \beq
    \#(i) = \left\{ \begin{array}{ll}
    0, & \hbox{if $A_i$ is a gluon;} \\
    1, & \hbox{if $A_i$, is a ghost} \\
    -1 & \hbox{if $A_i$, is an anti-ghost} \\
    \end{array} \right\}.
    \label{e-def-ghost-mumber}
    \eeq
If $t^\alpha$ are hermitian generators of the Lie algebra of $U(N)$, then
$A_i = A_i^\alpha t^\alpha$ with $A_i^{\alpha *} = A_i^\alpha$. For a gluon, $\#(i) =0$ and $A_i^\alpha$ is a real number while for a ghost or anti-ghost, $\#(i) = \pm 1$ and $A_i^\alpha$ is a real grassmann variable. Moreover,
    \beq
    (A_i^\a A_j^\b)^* = (-1)^{\#(i)\#(j)}  A_j^\b A_i^\a.
    \label{e-complex-conj}
    \eeq
Note that the ghost matrices are not related to the anti-ghost matrices by hermiticity.

It is also useful to define the ghost number of a tensor $C^{i_1
\cdots i_n}$ to be that of the multi-index: $\#(i_1 \cdots i_n) =
\sum_{k=1}^n \#(i_k)$ \footnote{For a general tensor $G = G_I \xi^I$,
$\#(G) = n$ if $\#(I) = n$ for each $I$ for which $G_I$
is non-vanishing. A general tensor has a well-defined ghost
number only if all terms have the same ghost number. Capitals denote multi-indices $I= i_1 i_2 \cdots i_n$ and repeated indices are summed.}. In keeping
with the structure of (\ref{e-gauge-fixed-ym-action}) we allow
gluon, ghost and anti-ghost matrices in the quadratic and cubic
terms of (\ref{e-action-ym-mat-mod-with-ghosts}) but only equal
numbers of ghosts and anti-ghosts in any term. In the cubic term,
$[A_j,A_k]$ denotes the anti-commutator if neither $A_j$ nor $A_k$
is a gluon and the commutator otherwise. In the quartic term of
(\ref{e-action-ym-mat-mod-with-ghosts}) we allow only gluons. In
other words, we assume $g^{ij}$ vanishes if either $i$ or $j$
corresponds to a ghost or anti-ghost index and that the ghost
numbers of $C^{ij}$ and $C^{ijk}$ vanish.
(\ref{e-gauge-fixed-ym-action}) can be regarded as a limiting case
of (\ref{e-action-ym-mat-mod-with-ghosts}) for appropriate integral
kernels $C^{ij}, C^{ijk}$ and  $g^{ij}$, when the indices become
continuous.

The action of our matrix model will be written as $\tr S(A) = \tr
S^I A_I$. It defines coupling tensors $S^I$. The partition function
is defined as $Z = \int \Pi_j dA_j e^{-N \tr S(A)}$ where the
integration is over all independent matrix elements of the gluon,
ghost and anti-ghost matrices. Observables are correlation tensors
    \beq
    \bra \Ntr A_{I_1} \cdots \Ntr A_{I_n} \ket =
    \ov{Z} \int \Pi_j dA_j  e^{-N \tr S(A)} \Ntr
    A_{I_1} \cdots \Ntr A_{I_n}.
    \label{e-multi-trace-correlations}
    \eeq
They are symmetric in the $I_k$'s up to a possible sign. For example
    \beq
    \bra \Ntr A_{I} \Ntr A_{J} \ket
    = (-1)^{\#(I) \#(J)} \bra \Ntr A_{J} \Ntr A_{I} \ket.
    \eeq
This is because matrix elements $[A_i]^a_b$ are graded
commutative. So we pick up a minus sign under transposition of ghost
or anti-ghost matrices
    \beq
    [A_i]^a_b [A_j]^c_d = (-1)^{\#(i)
    \#(j)} [A_j]^c_d [A_i]^a_b.
    \eeq
It follows that the trace operation is graded cyclic,
for example $\tr A_i A_j A_k = (-1)^{\#(k) \#(ij)} \tr A_k A_i A_j$.

We restrict to actions $S^I A_I$ whose non-vanishing coupling
tensors have zero ghost number. Gauge-fixed Yang Mills theory
(\ref{e-gauge-fixed-ym-action}) is of this type. In such a theory,
correlations of a tensor with non-zero total ghost number must
vanish\footnote{Consequence of Feynman rules: interaction vertices
have equal numbers of ghost and anti-ghost lines attached.}
    \beq
    \bra\Ntr A_{I_1} \cdots \Ntr A_{I_n} \ket = 0 {\rm ~~if~~} \#(I_1) + \#(I_2) + \cdots + \#(I_n) \ne 0.
    \eeq
For example $\bra \ov{N} \tr A_g A_g A_c A_c \ket =0$ but $\bra \ov{N}
\tr A_g A_c A_{\bar c} \ket$ can be non-trivial, where $g,c$ and
$\bar c$ stand for gluon, ghost and anti-ghost. Multi-trace
correlators factorise into single trace correlations in the large-$N$ limit:
     \beq
    G_{I} = \lim_{N \to \infty} \bra\Ntr A_{I} \ket, {\rm ~~and~~}
    \bra\Ntr A_{I_1} \cdots \Ntr A_{I_n} \ket &=& G_{I_1} \cdots G_{I_n}
    + {\cal O}(1/N^2).
     \eeq
The hermiticity (\ref{e-complex-conj}) of the matrices in colour space implies an order reversal property under complex conjugation
$G_{i_1 i_2 \cdots i_n}^* = (-1)^p G_{i_n i_{n-1} \cdots i_2 i_1}$ where
$p  = \#(i_1) \#(i_2 \cdots i_n) + \#(i_2) \#(i_3 \cdots i_n) + \cdots + \#(i_{n-1}) \#(i_n)$.
The factorised correlation tensors $G_I$ are in general graded
cyclic:
    \beq
    G_{Ii} = (-1)^{\#(i)\#(I)} G_{iI}, ~~~~
    G_{IJ} = (-1)^{\#(I)\#(J)} G_{JI}.
    \eeq
Similarly, the only part of $S^I$ that contributes is its graded
cyclic projection
    \beq
        S^{Ii} \mapsto \ov{|Ii|}[S^{Ii} + S^{iI} (-1)^{\#(I)\#(i)}
         + \cdots].
    \eeq
So we assume that $S^{IJ} = S^{JI} (-1)^{\#(I) \#(J)}$ for all
$I,J$, i.e. that $S^I$ is graded cyclic.

%---------------------------------------
\section{Schwinger-Dyson to loop equations in the large-$N$ limit}
\label{s-sde-to-loop-eq}
%---------------------------------------

The loop equations are quantum corrected equations of motion for correlation tensors.
To derive them, we consider changes of variable (vector fields)
    \beq
    A_i \mapsto A_i^\pr = A_i + v^I_i A_I.
    \label{e-change-of-variable}
    \eeq
Here $v^I_i$ could either be a small real number or a grassmann number
with ghost number $\#(v^I_i) = \pm 1$. However, we cannot change a
complex number by a grassmann-valued quantity and vice versa, so we
require that $v^I_i$ are such that
    \beq
    \#(i) = \#(v^I_i) + \#(I).
    \eeq
An example of such a change of variable is a BRST transformation in
Yang-Mills theory
    \beq
    A_\mu^\a \mapsto A_\mu^\a - \ov{g} (D_\mu c^\a) \la; ~~~
    c^\a \mapsto c^\a - \half f^{\a \b \gamma} c^\b c^\gamma \la; ~~~
    \bar c^\a \mapsto \bar c^\a - \ov{\xi g} (\pdr^\mu A_\mu^\a) \la.
    \eeq
Here $v^I_i \propto \la$ is a constant grassmann quantity with ghost
number $-1$, i.e. $\#(v^I_i) = -1$ and it is easily seen that the
conditions $\#(i) = \#(v^I_i) + \#(I)$ are satisfied for the BRST
transformation.

Under (\ref{e-change-of-variable}), $A_K \mapsto A_K + \d_K^{LiM}
v_i^I A_{L I M}$. So if we define the `loop' variable $\Phi_K = \N \tr A_K$,
    \beq
    \F_K &\mapsto& \F_K + \d_K^{L i M } v_i^I \F_{L I M} \cr
    \F_{K_1} \cdots \F_{K_n} &\mapsto& \F_{K_1} \cdots \F_{K_n}
        + \sum_{p=1}^n \d_{K_p}^{L_p i M_p } v_i^I \F_{L_p I M_p} \cr
    e^{-N \tr S^J A_J} &\mapsto& e^{-N \tr S^J A_J} \bigg[1
    - N^2 v^I_i S^{J_1 i J_2} \F_{J_1 I J_2} + \cdots
        \bigg] \cr
    \det{\bigg( \dd{A_i^\pr}{A_j}\bigg)} &=& \det\bigg(
        \dd{[A_i]^a_b + v_i^I [A_I]^a_b}{[A_j]^c_d}\bigg) \cr
    &=& \det(\d^j_i \d^a_c \d^d_b + v^I_i (-1)^{\#(j) \#(I_1)} \d_I^{I_1 j I_2}
    [A_{I_1}]^a_c [A_{I_2}]^d_b) \cr
    &=& 1 + N^2 v_i^I \d^{I_1 i I_2}_I (-1)^{\#(i) \#(I_1)} \F_{I_1} \F_{I_2} + \cdots.
    \eeq
The sign in the Jacobian comes from moving the grassmann (left)
derivative through the product. Requiring the invariance of
(\ref{e-multi-trace-correlations}) under the changes of integration
variables (\ref{e-change-of-variable}) leads to the finite-$N$
Schwinger-Dyson equations (SDE)
    \beq
    v^I_i S^{J_1 i J_2} \bra \F_{J_1 I J_2} \ket = v^{I_1 i I_2}_i
    (-1)^{\#(i) \#(I_1)} \bra \F_{I_1} \F_{I_2} \ket
     + \ov{N^2} \sum_{p=1}^n \d_{K_p}^{L_p i M_p} v^I_i \bra \F_{L_p I
     M_p} \ket.
    \eeq
In the large-$N$ limit, the factorised SDE ignoring the last term
are
    \beq
    v^I_i S^{J_1 i J_2} G_{J_1 I J_2}  = v^{I_1 i I_2}_i
    (-1)^{\#(i) \#(I_1)} G_{I_1} G_{I_2}.
    \eeq
Since correlations $G_{I_1}$ vanish if $\#(I_1) \ne 0$ we get
    \beq
    v^I_i S^{J_1 i J_2} G_{J_1 I J_2}  = v^{I_1 i I_2}_i
    G_{I_1} G_{I_2} ~~~ {\rm for ~~all~~ vector ~~fields~~} v.
    \eeq
Now taking $v^I_i$ to be non-vanishing only for a fixed $i,I$ gives
us the loop equations (LE)
    \beq
    S^{J_1 i J_2} G_{J_1 I J_2}  = \d^{I_1 i I_2}_I
    G_{I_1} G_{I_2}   ~~~ \forall ~~I,~i.
    \eeq
Using graded cyclicity of $S^I$ and $G_I$ we write this as ($|I|$ is the length of the multi-index $I$)
    \beq
    |Ji| S^{Ji} G_{JI} = \d_I^{I_1 i I_2} G_{I_1} G_{I_2}.
    \eeq
The loop equations have the same form as for a bosonic matrix model (see \cite{deform-prod-der}).

%---------------------------------------------
\section{Loop equations in terms of concatenation and left annihilation}
\label{s-LE-conc-left-ann}
%---------------------------------------------

Define the left annihilation operator acting on the generating
series of correlations $G(\xi) = G_I \xi^I$
    \beq
    [D_j G]_I = G_{jI} {~~ \rm and ~~}
    [D_{j_n} D_{j_{n-1}} \cdots D_{j_1} G]_I  = [D_{j_n \cdots j_1} G]_I = G_{j_1 \cdots j_n I}.
    \label{e-left-annihilation}
    \eeq
The ghost number $\#(D_j G) = \#(j)$ if $G(\xi)$ has zero ghost number. In terms of the concatenation product
    \beq
    [F G]_I = \d_I^{JK} F_J G_K;~~~~ F(\xi) G(\xi) = F_I G_J \xi^{IJ},
    \eeq
the LE take the same form as in the purely bosonic case \cite{deform-prod-der}:
 \beq
    \sum_{n \geq 0} (n+1) S^{j_1 \cdots j_n i} D_{j_n \cdots j_1} G(\xi)
        &=& G(\xi)\xi^i G(\xi),
    {\rm ~~~or~~~} {\cal S}^i G(\xi) = G(\xi)\xi^i G(\xi)
    \cr {\rm where~~}  {\cal S}^i &=& \sum_{n \geq 0} (n+1) S^{j_1 \cdots j_n i} D_{j_n \cdots j_1}.
    \label{e-ghost-loop-eqns}
 \eeq
Both LHS and RHS have ghost number $\#(i)$.

{\flushleft \bf Graded commutator} To simplify notation below, we
introduce the graded commutator
    \beq
    [D_i,D_j] = D_i D_j - (-1)^{\#(i) \#(j)} D_j D_i.
    \label{e-graded-commutator}
    \eeq
It is graded symmetric $[D_i,D_j] = -(-1)^{\#(i) \#(j)} [D_j,D_i]$,
and reduces to the commutator if either $i$ or $j$ is a gluon and to
the anti-commutator if neither is a gluon. We use the same notation
for commutators and anti-commutators of the $A_i$.

%---------------------------
\section{Example: Ghost terms in Yang-Mills action}
\label{s-ghost-mat-model}
%---------------------------

As an example, consider a class of matrix models inspired by the
ghost terms in the gauge-fixed Yang-Mills
action(\ref{e-gauge-fixed-ym-action}) $\pdr_\mu \bar c ~ \pdr^\mu c
- ig~ \pdr_\mu \bar c ~[A^\mu,c]$. The action is
    \beq
    \tr S = \half \tr C^{ij} A_i A_j + \tr C^{ijk} A_i [A_j, A_k]
    \label{e-action-ghost-mat-model}
    \eeq
where $[,]$ is the graded commutator (\ref{e-graded-commutator}). If all the matrices
are gluons, then this is a zero momentum Gaussian +
Chern-Simons-type matrix model. Here we allow gluon, ghost and
anti-ghost matrices so that for special choices of $C^{ij}$ and
$C^{ijk}$, this action can also model the terms $\pdr_\mu \bar c ~
\pdr^\mu c$ and $i g \pdr_\mu \bar c ~[A^\mu,c]$ appearing in the
gauge-fixed Yang-Mills action. We obtain the coupling tensors
$S^{ij}$ and $S^{ijk}$, which can be chosen to be
graded-cyclic. Then we obtain the factorised LE and the differential
operator ${\cal S}^i$. First, write the action as
    \beq
    \tr S &=& \half \tr C^{ij} A_i A_j + \tr C^{ijk} A_i \{ A_j A_k
        - (-1)^{\#(j) \#(k)} A_k A_j \} \cr
    &=&  \half \tr C^{ij} A_i A_j + \tr \bigg[ C^{ijk} A_{ijk} - (-1)^{\#(j) \#(k)} C^{ikj}
        A_{ijk} \bigg].
    \eeq
Thus the coupling tensors are
    \beq
    S^{ij} = \half C^{ij} {\rm ~~~ and ~~~}
    S^{ijk} = C^{ijk} - (-1)^{\#(j) \#(k)} C^{ikj}.
    \label{e-coupling-tensors-CS-ghost}
    \eeq
From graded cyclicity of the trace, it follows that the coupling tensors are
graded cyclic
    \beq
    S^{ij} = (-1)^{\#(i) \#(j)} S^{ji} {\rm ~~~and ~~~}
    S^{ijk} = (-1)^{\#(k) \#(ij)} S^{kij}.
    \eeq
As for the differential operator ${\cal S}^i$, from (\ref{e-ghost-loop-eqns}) we have
    \beq
    {\cal S}^i = C^{ji} D_j + 3 S^{jki} D_{kj}
        = C^{ji} D_j + 3 \bigg(C^{jki} -
        (-1)^{\#(k) \#(i)} C^{jik} \bigg) D_k D_j.
    \eeq
The interesting question (which we will answer in the affirmative in section \ref{s-var-of-action-is-derivation}) is whether ${\cal S}^i$ is a derivation of the graded shuffle product of ghost number zero
tensors. Mere graded cyclicity of $S^{ijk}$ is not sufficient for
this. Rather, it is useful to write ${\cal S}^i$ in terms of graded commutators of left annihilations. The reason is that the left annihilation $D_i$ is like a first order differential operator (vector field) when acting on the shuffle algebra. While products of vector fields $D_i D_j$ are no longer vector fields, their commutators $[D_i, D_j]$ continue to be vector fields and therefore define derivations of the shuffle algebra.

In terms of the graded commutator (\ref{e-graded-commutator}),
the action may be written as
    \beq
    \tr S = \tr \half C^{ij} A_i A_j + \tr C^{ijk} A_i [A_j,A_k]
        = \tr \half C^{ij} A_i A_j + \tr C^{ijk} [A_i,A_j] A_k.
    \eeq
Graded symmetry of the graded commutator in turn implies graded
symmetry of $C^{ijk}$
    \beq
    C^{ijk} + (-1)^{\#(j) \#(k)} C^{ikj} =0
    {\rm ~~~and~~~} C^{ijk} + (-1)^{\#(i) \#(j)} C^{jik} =0.
    \eeq
More precisely, even if these quantities were non-vanishing, they
would not contribute to the action. Therefore they can be taken to
vanish without loss of generality. Using this property twice we can
write
    \beq
    {\cal S}^i &=& C^{ji} D_j +
        3 C^{jki} D_k D_j - 3(-1)^{\#(j) \#(i)} C^{kij} D_j D_k \cr
    &=& C^{ji} D_j + 3C^{jki} D_k D_j - 3(-1)^{\#(k) \#(j)} C^{jki} D_j D_k
    ~=~ C^{ji} D_j  + 3 C^{ijk} [D_k,D_j].
    \eeq
Thus ${\cal S}^i$ is a linear combination of the left annihilation
and its (anti-)commutators. Finally, the factorised LE for the
action (\ref{e-action-ghost-mat-model}) are
    \beq
    C^{ji} D_j G(\xi) + 3 C^{ijk} [D_k,D_j] G(\xi) = G(\xi) \xi^i
    G(\xi).
    \eeq
More generally, ${\cal S}^i$ for the full gauge-fixed Yang-Mills
action is a linear sum of the following combinations of left
annihilation operators
 \beq
    D_g,~ D_c,~ D_{\bar c},~ [D_c,D_{\bar c}]_+~,~~ [D_c, D_g]_-
    ~,~~ [D_{\bar c},D_g]_-~,~ ~ [D_{g_1}, D_{g_2}]_-
    ~,~~ [D_{g_1},[D_{g_2},D_{g_3}]_-]_-.
 \eeq
Each of these combinations arises as the variation of one or more
terms in the gauge-fixed Yang-Mills action
(\ref{e-gauge-fixed-ym-action}). $D_g$ comes from varying terms
involving two derivatives and two gluon fields (e.g. $\ov{2 \xi}
(\pdr^\mu A_\mu)^2$ and $\pdr_\mu A_\nu(\pdr^\nu A^\mu ~-~ \pdr^\mu
A^\nu)$). $D_c$ and $D_{\bar c}$ come from $\pdr_\mu \bar c
~\pdr^\mu c$. The anti-commutator $[D_c,D_{\bar c}]_+$ arises from
the variation of $g \pdr_\mu \bar c ~[A^\mu,c]$ with respect to the
gluon. The commutators $[D_c, D_g]$ and $[D_{\bar c},D_g]$ arise
from varying the same term with respect to an anti-ghost or a ghost
(as in the example above). $[D_{g_1}, D_{g_2}]$ originates from
varying the term linear in derivatives: $g \pdr_\mu A_\nu
[A^\mu,A^\nu]$. Finally $[D_{g_1},[D_{g_2},D_{g_3}]]$ has its origin
in the term independent of momentum ${g^2 \over 4}
[A_\mu,A_\nu][A^\mu,A^\nu]$.

We showed in \cite{deform-prod-der} that ${\cal S}^i$ for the purely
gluonic part of the Yang-Mills + Chern-Simons + Gaussian action is a
derivation of the shuffle product of gluon correlations. Here we
show that even when gauge fixing and ghost terms are included, ${\cal
S}^i$ is a derivation of the graded shuffle product of gluon-ghost
correlations.

%---------------------------------------------
\section{Graded shuffle product}
\label{s-graded-shuffle}
%---------------------------------------------

We will call the extension of the shuffle product (see \cite{deform-prod-der,reutenauer}) to correlation
tensors of gluon and ghost matrices by the name graded shuffle
product. It is essentially the shuffle product with minus signs when
ghost or anti-ghost indices are transposed. The definition is
    \beq
    [F \circ G]_I
    \equiv \sum_{I_1 \sqcup I_2 = I} (-1)^{\gamma(I;I_1,I_2)}
        F_{I_1} G_{I_2}.
    \label{e-graded-shuffle-product}
    \eeq
$I_1 \sqcup I_2 = I$ is the condition that $I_1$ and $I_2$ are
complementary order preserving sub-strings of $I$. In other words, we riffle-shuffle the card packs $I_1$ and $I_2$.
For example if
$I= i_1 i_2 i_3 i_4$ then one permissible choice is $I_1 = i_2 i_4$
and $I_2 = i_1 i_3$ while $I_1 = i_3 i_2$, $I_2 = i_1 i_4$ is not allowed.
We call $\gamma(I;I_1,I_2)$ the {\it ghost
crossing number} of the ordered triple $(I; I_1, I_2)$. It is just
zero for bosonic matrix models. More generally, the string $I$ is
transformed into the string $I_1 I_2$ by a minimum number of
transpositions of neighboring indices. Each transposition $i_p i_q
\mapsto i_q i_p$ contributes $\#(i_p) \#(i_q)$. $\#(i_p) \#(i_q)$ is
$0$ if both $i_p, i_q$ are gluons, $+1$ if both are ghosts or
anti-ghosts and $-1$ if one was a ghost and the other an anti-ghost.
The sum of these contributions is the ghost crossing number. For
example let
 \beq
    I = i_1 i_2 i_3 i_4 i_5; ~~~~
    I_1 = i_1 i_4 i_5;~~~ I_2 = i_2 i_3.
 \eeq
The sequence of transpositions may be
 \beq
    i_1 i_2 i_3 i_4 i_5  &\rightarrow&
    i_1 i_2 i_4 i_3 i_5  (-1)^{\#(i_4) \#(i_3)} \rightarrow
    i_1 i_4 i_2 i_3 i_5 (-1)^{\#(i_4)\#(i_3) + \#(i_4)\#(i_2)} \cr
    &\rightarrow&
    i_1 i_4 i_2 i_5 i_3 (-1)^{\#(i_4)\#(i_3) + \#(i_4)\#(i_2)+ \#(i_5)\#(i_3)}
    \cr &\rightarrow& i_1 i_4 i_5 i_2 i_3 (-1)^{\#(i_4)\#(i_3) + \#(i_4)\#(i_2)+
        \#(i_5)\#(i_3)+ \#(i_5)\#(i_2)}
    \cr &\rightarrow& i_1 i_4 i_5 i_2 i_3 (-1)^{\#(i_4 i_5)\#(i_2
    i_3)}.
 \eeq
The {\it ghost crossing sign} $(-1)^{\gamma(I;I_1,I_2)}$ is
independent of the choice of sequence of transpositions, so we don't
need to stick to the minimum number to find the sign. Moreover,
    \beq
    (-1)^{\gamma(I;J,K)} = (-1)^{\gamma(I;K,J) + \#(J) \#(K)}.
    \eeq
In the sequel we will be interested in tensors $G_I$ of zero ghost
number, since the others vanish.

{\flushleft \bf Preservation of zero ghost number:} The graded
shuffle product of two tensors of zero ghost number is again a
tensor of zero ghost number. Suppose $F_{I_1}$ and $G_{I_2}$ have
zero ghost numbers, $\#(I_1) = \#(I_2) = 0$. From
(\ref{e-graded-shuffle-product}) and (\ref{e-def-ghost-mumber}), if $I =
I_1 \sqcup I_2$, then $\#(I) = \#(I_1) + \#(I_2) = 0$. So $(F \circ
G)_I$ has zero ghost number.

{\flushleft \bf Commutativity of graded shuffle:} The graded shuffle
product $F \circ G$ is commutative if either $F$ or $G$ has zero
ghost number:
 \beq
        [F \circ G]_I &=& \sum_{I_1 \sqcup I_2 = I}
            (-1)^{\gamma(I;I_1,I_2)} F_{I_1} G_{I_2}
        = \sum_{I_1 \sqcup I_2 = I} (-1)^{\gamma(I;I_2,
            I_1) + \#(I_1) \#({I_2})} F_{{I_1}} G_{{I_2}} \cr
        &=& \sum_{I_2 \sqcup I_1 = I} (-1)^{\gamma(I;I_1,
            I_2) + \#({I_2}) \#({I_1})} F_{{I_2}} G_{{I_1}}
        = \sum_{I_1 \sqcup I_2 = I} (-1)^{\gamma(I;I_1,
            I_2)} G_{{I_1}} F_{{I_2}}
        = [G \circ F]_I.
 \eeq

%---------------------------------------------
\section{Left annihilation is a derivation of graded shuffle product}
\label{s-left-ann-derivation-graded-sh-prod}
%---------------------------------------------

$D_i$ is a derivation of the graded shuffle product of two ghost
number zero tensors: $D_i (F \circ G) = (D_i F) \circ G + F \circ
(D_i G) ~~ {\rm if~~} \#(F) = \#(G) = 0.$ To show this, we write
 \beq
    [D_i (F \circ G)]_I = [F \circ G]_{iI}
        = \sum_{I_1 \sqcup I_2 = iI} (-1)^{\gamma(iI;I_1,I_2)}
            F_{{I_1}} G_{{I_2}}.
 \eeq
Now either $i \in I_1$ or $i \in I_2$, so
 \beq
    [D_i (F \circ G)]_I &=& \sum_{I_1 \sqcup I_2 = I} \bigg[
        (-1)^{\gamma(iI;iI_1,I_2)} F_{i{I_1}}
        G_{{I_2}} + (-1)^{\gamma(iI;{I_1},i {I_2})}
        F_{{I_1}} G_{i {I_2}} \bigg]  \cr
    &=& \sum_{I_1 \sqcup I_2 = I} \bigg[ (-1)^{\gamma(iI;i {I_1},{I_2})} [D_i F]_{{I_1}}
        G_{{I_2}} + (-1)^{\gamma(iI;{I_1},i {I_2})}
        F_{{I_1}} [D_i G]_{{I_2}} \bigg].
 \eeq
Since $i$ doesn't cross any index, $(-1)^{\gamma(iI; i {I_1},{I_2})}
= (-1)^{\gamma(I;{I_1},{I_2})}$. Similarly, $(-1)^{\gamma(iI;{I_1},i
{I_2})}=(-1)^{\#(i)\#(I_1) + \gamma(I;{I_1},{I_2})}$. Thus
 \beq
    [D_i (F \circ G)]_I = \sum_{I_1 \sqcup I_2 = I} \bigg[
        (-1)^{\gamma(I;{I_1},{I_2})} [D_i
        F]_{{I_1}} G_{{I_2}} + (-1)^{\#(i)\#({I_1})
        + \gamma(I;{I_1},{I_2})} F_{{I_1}} [D_i G]_{{I_2}}  \bigg].
 \eeq
Since correlation tensors vanish for non-zero ghost number we can
take $\#({I_1}) =0$. Hence $[D_i (F \circ G)]_I = [(D_i F) \circ
G]_I + [F \circ (D_i G)]_I.$ Thus $D_i$ is a derivation of the
shuffle product of two tensors provided each has zero ghost number.
More generally, if no assumption is made on the ghost number of $F$
and $G$, then
    \beq
    [D_i (F \circ G)]_I = [(D_i F) \circ G]_I
        + \sum_{I_1 \sqcup I_2 = I} (-1)^{\#(i)\#({I_1})
        + \gamma(I;{I_1},{I_2})} F_{{I_1}} [D_i G]_{{I_2}}.
    \label{e-more-gen-left-ann-of-shuff-prod}
    \eeq
{\flushleft \bf Graded commutator of left annihilation:} In the
bosonic theory, a commutator $[D_i,D_j]$ of derivations is a
derivation of the shuffle product \cite{deform-prod-der}. More generally, we will show that if $F, G$ have
zero ghost number, then the graded commutator of left annihilations is a derivation of their shuffle product:
    \beq
    [D_i, D_j](F \circ G) = ([D_i, D_j] F) \circ G
        + F \circ ([D_i , D_j] G).
    \eeq
To show this, use the derivation property of $D_j$ and then the more
general result (\ref{e-more-gen-left-ann-of-shuff-prod}) for $D_i$.
    \beq
    D_i D_j (F \circ G) &=& D_i (D_j F \circ G)
            + D_i(F \circ D_j G) \cr
    [D_i D_j (F \circ G)]_I &=& [D_{ij} F \circ
        G]_I + [D_i F \circ D_j G]_I
     + \sum_{I_1 \sqcup I_2 = I} (-1)^{\#(i) \#({I_1}) + \gamma(I; {I_1},
        {I_2})} \cr && \times [D_j F]_{{I_1}} [D_i G]_{{I_2}}
    + \sum_{I_1 \sqcup I_2 = I} (-1)^{\#(i) \#({I_1}) + \gamma(I; {I_1},
        {I_2})} F_{{I_1}} [D_{ij} G]_{{I_2}}.
    \eeq
In the third term $\#({I_1}) = - \#(j)$ and in the last term,
$\#({I_1}) = 0$, thus
    \beq
    D_{ij}(F \circ G) &=& D_{ij} F \circ G +
        F \circ D_{ij} G + D_i F \circ D_j G +
        (-1)^{\#(i)\#(j)} D_j F \circ D_i G \cr
    D_{ji}(F \circ G) &=& D_{ji} F \circ G +
        F \circ D_{ji} G + D_j F \circ D_i G +
        (-1)^{\#(i)\#(j)} D_i F \circ D_j G.
    \eeq
Combining these we find that the graded commutator of left
annihilations is a derivation
    \beq
    [D_i, D_j](F \circ G) &=& D_i D_j (F \circ G) - (-1)^{\#(i) \#(j)}
        D_j D_i (F \circ G) \cr
    &=& (D_{ij} F - (-1)^{\#(i) \#(j)} D_{ji} F) \circ G +
        F \circ (D_{ij}G - (-1)^{\#(i) \#(j)} D_{ji} G) \cr
    &=& ([D_i,D_j]F) \circ G + F \circ ([D_i, D_j] G).
    \eeq
{\flushleft \bf Iterated commutator of gluonic left annihilation:}
If $i,j,k$ all have zero ghost numbers, then $[[D_i,D_j],D_k]$ is a
derivation of the shuffle product of two tensors of zero ghost
number each. This is a consequence of the derivation property of
$D_k$ and $[D_i,D_j]$ for shuffle products of tensors of zero ghost
number.
    \beq
    [[D_i,D_j],D_k] (F \circ G) &=& [D_i,D_j](D_k F \circ G + F \circ D_k G)
        \cr && - D_k([D_i,D_j]F \circ G) - D_k (F \circ [D_i,D_j] G)
    \eeq
Here each of the terms within parentheses is shuffle product of
tensors of zero ghost number since $i,j,k,F,G$ are. Applying the
derivation property of $D_k$ and $[D_i,D_j]$ again, four of the
terms cancel out and we get the desired result
    \beq
    [[D_i,D_j],D_k] (F \circ G) = [[D_i,D_j],D_k] F \circ G + F
    \circ [[D_i,D_j],D_k] G.
    \eeq

%------------------------------------------------------------------
\section{${\cal S}^i$ for Yang-Mills matrix model
with ghosts is a derivation of shuffle algebra}
\label{s-var-of-action-is-derivation}
%------------------------------------------------------------------

Let us now consider the Yang-Mills matrix model with ghosts
introduced in (\ref{e-action-ym-mat-mod-with-ghosts})
    \beq
    S = \half \tr C^{ij} A_i A_j + \tr C^{ijk} A_i [A_j, A_k]
    -\ov{4 \a} \tr [A_i,A_j][A_k,A_l] g^{ik} g^{jl}
    \eeq
This differs from the model considered in
(\ref{e-action-ghost-mat-model}) by the addition of the quartic term,
which however involves only gluons. The latter was studied in
\cite{deform-prod-der}. Using our results from
\cite{deform-prod-der} and section \ref{s-ghost-mat-model} we get
the loop equations ${\cal S}^i G(\xi) = G(\xi) \xi^i G(\xi)$ where
${\cal S}^i$ is
    \beq
    {\cal S}^i = C^{ji} D_j  + 3 C^{ijk} [D_k,D_j] -\ov{\a} g^{ik}
    g^{jl} [D_j[D_k,D_l]].
    \eeq
${\cal S}^i$ is a linear combination of $D_j$, its graded
commutators and gluonic iterated commutator $[D_j[D_k,D_l]]$. Each
of these was shown to be a derivation of the graded shuffle product
in section \ref{s-left-ann-derivation-graded-sh-prod}. We conclude
that the Schwinger-Dyson operator ${\cal S}^i$ for Yang-Mills matrix models with ghosts is a derivation of the graded shuffle product of ghost number zero correlation tensors. Based on the bosonic case in \cite{deform-prod-der}, one might have suspected that $D_i$ and ${\cal S}^i$ would, at best, be graded derivations upon including ghosts (i.e. satisfy the Leibnitz rule up to a sign). But they turn out to be ordinary derivations of the graded shuffle product, since correlations with non-zero ghost number vanish in the large-$N$ limit.

%------------------------------------------------------------------
\section{Discussion}
\label{s-discussion}
%------------------------------------------------------------------

Physically, the Schwinger-Dyson operator ${\cal S}^i$ arises from the variation of the action and is therefore a classical ($\hbar =0$) concept. Indeed, the correlations in the limit $\hbar \to 0$ are annihilated by the Schwinger-Dyson operator. For Yang-Mills matrix models with ghosts, ${\cal S}^i = C^{ji} D_j  + 3 C^{ijk} [D_k,D_j] -\ov{\a} g^{ik} g^{jl} [D_j[D_k,D_l]]$ is a linear combination of iterated (anti) commutators of the left annihilation operator. The quantum effects in the large-$N$ limit are encoded in the variation of the measure, or the quadratic term in the loop equations ${\cal S}^i G(\xi) = G(\xi) \xi^i G(\xi)$. $\xi^i$ are external sources and $G(\xi)$ is the generating series of correlations, the product on the rhs is concatenation. $i,j,k$ label the matrices and in the continuum limit would label space-time points. We have shown that there is a mismatch between the `classical' Schwinger-Dyson operator and the `quantum' concatenation product, so that ${\cal S}^i$ is not a differential operator in the loop equations. This `mismatch' makes the equations both interesting and hard to solve. However, we identified a new commutative shuffle product, with respect to which ${\cal S}^i$ is indeed a differential operator. Shuffle product is the point-wise product of Wilson loops, written in terms of correlations. As suggested in \cite{deform-prod-der}, we can expand concatenation around shuffle to exploit the derivation property of left annihilation $D_i$. At $0^{\rm th}$ order, the loop equations become differential equations in the graded-shuffle algebra ${\cal S}^i G(\xi) = G(\xi) \circ \xi^i \circ G(\xi)$. Moreover, we showed that ${\cal S}^i$ is not just a differential operator with respect to shuffle, but behaves as a {\rm first order} differential operator. This derivation property of ${\cal S}^i$ leads to a further simplification. First we define the reciprocal of $G(\xi)$ with respect to the graded shuffle product. Since $G_0=1$ is non-vanishing, $G(\xi)$ has a right reciprocal $G(\xi) \circ F(\xi) =1$. Moreover, non-vanishing $G_I$ have zero ghost number, so the same is true of the $F_I$. Thus, the graded shuffle product $G \circ F$ is commutative and the reciprocal is unique and two-sided. Explicitly, $F_0 = 1$, $F_i = - G_i$, $F_{ij} = - G_{ij} +  G_i G_j \{1 + (-1)^{\#(i) \#(j)} \}$ etc. More generally, for $|I| \geq 1$,
    \beq
    F_I = - G_I - \sum_{\stackrel{I_1 \sqcup I_2 =I}{I_1 \ne I, I_2 \ne I}} (-1)^{\gamma(I,I_1,I_2)} F_{I_1} G_{I_2}
    \eeq
expresses $F_I$ in terms of $G_K$'s and lower order $F_J$'s. Iterating, we can find the shuffle reciprocal.
Now, since ${\cal S}^i$ satisfies the Leibnitz rule and ${\cal S}^i(1)=0$,
    \beq
    0 = {\cal S}^i (G(\xi) \circ F(\xi)) = {\cal S}^i G \circ F + G \circ {\cal S}^i F ~~\implies~~
    {\cal S}^i G = - G \circ {\cal S}^i F \circ G.
    \eeq
Thus the loop equations become linear equations for the graded shuffle reciprocal
    \beq
    {\cal S}^i F = - \xi^i.
    \eeq
This substantial simplification is {\em absent} for a generic matrix model whose Schwinger-Dyson operator is not a derivation of the graded shuffle product. This underscores the potential practical importance of the derivation property of the Schwinger-Dyson operator of Yang-Mills matrix models, in a possible approximation method based on expanding concatenation around shuffle.

We can give another physical interpretation. The large-$N$ limit is a `classical' limit since $U(N)$ invariants stop fluctuating in this limit, though $\hbar =1$ is held fixed. However, even this large-$N$ `classical' limit is difficult to solve, partly because we have non-commutative concatenation products left over. Replacing concatenation by commutative shuffle products may be thought of as taking a further classical limit. Indeed, when this is done, we found that the equations become linear! However, much work still needs to be done. In particular, for some models, we found that the loop equations are under-determined \cite{deform-prod-der}. In those cases, the above equations have to be supplemented by non-anomalous Ward identities arising from Schwinger-Dyson equations that are naively $1/N^2$ suppressed \cite{non-anom-ward-id}. We hope to report on more explicit calculations in a future publication.

Mathematically, together with \cite{deform-prod-der}, our work shows that it may be fruitful to think of the loop equations as living in the differential bi-algebra formed by concatenation, shuffle and their derivations. Roughly, the space of based oriented loops (modulo backtracking) on space-time, $Loop(M)$, is to be regarded as a free group on a continuously infinite number of generators labeled by the loops $\gamma$. The concatenation of loops and reversal of loops are the product and inverse operations in this gigantic free group. A typical function on $Loop(M)$ is a Wilson loop expectation value with respect to a $U(N)$ connection. The space of functions on this $Loop(M)$ is then automatically a commutative Hopf algebra under the commutative point-wise product of functions $(FG)(\gamma) = F(\g) G(\g)$ and comultiplication given by concatenation of loops: $(\D'G)(\g_1,\g_2) = G(\g_1 \g_2)$. Formally, we also have a dual co-commutative Hopf algebra defined via the group algebra of the free group of loops. The Makeenko-Migdal loop equations of large-$N$ Yang-Mills theory are defined in this space, using the path derivative of the area derivative (the analogue of our Schwinger-Dyson operator), which is a derivation of the pointwise product.

Hermitian multi-matrix models, regarded as discrete toy-models for gauge theory in physics, also provide a finitely generated toy-model for the above mathematical theory on loop space. The shuffle product plays the role of the commutative product of functions on $Loop(M)$. It is obtained by expanding the Wilson loop in iterated integrals of gluon correlations. The coproduct in the discrete model involves concatenation of tensors ($\D'(\xi^I) = \d^I_{JK} \xi^J \otimes \xi^K$) rather than loops. One difference (since we are dealing with hermitian rather than unitary matrices), is that we do not have inverses for the generators $\xi^i$ to play the role of reversal of loop orientation. As a consequence, the group algebra of the free group of loops is now replaced by the concatenation algebra of tensors (free associative algebra), which is the monoid algebra of the free monoid. The free associative algebra could also have been obtained as the universal envelope of the free Lie algebra. Moreover, left annihilation and its iterated commutators (including the physically relevant Schwinger-Dyson operator) form a free Lie algebra of derivations of the shuffle algebra. This suggests we should think of the Schwinger-Dyson operator of Yang-Mills theory as a vector field on the free group of loops on space time. Thus, despite being a discrete model, our setup preserves many of the algebraic and differential structures of the original theory on $Loop(M)$, which is difficult to study directly. We hope to investigate this correspondence in greater detail.

%----------------------------------
\section*{Acknowledgements}
%----------------------------------

We thank the EU for support in the form of a Marie Curie fellowship and the EPSRC for support via an EPSRC fellowship. We also thank S. G. Rajeev, A. Agarwal and L. Akant for discussions at an early stage of this work.

%------------------------------------

%------------------------------------

\end{document}